\documentclass[aps, prb, preprint, superscriptaddress]{revtex4-2}
\usepackage{silence}
\usepackage{graphicx}
\usepackage{makecell, multirow}
\usepackage{color, bm, soul, xcolor}
\usepackage{amsmath, amssymb}
\usepackage{geometry}
\usepackage{array,booktabs}
\usepackage{chemformula}
\usepackage{dcolumn}
\usepackage[colorlinks=true, linkcolor=blue, citecolor=blue, urlcolor=blue]{hyperref}

\newcommand{\note}[1]{{\textcolor{blue}{#1}}}
\newcommand{\Pca}{\ensuremath{\mathit{Pca}2_1}}
\newcommand{\Pc}{\ensuremath{\mathit{Pc}}}
\newcommand{\Cac}{\ensuremath{\mathit{C}_{ac}}}

\begin{document}

\title{Multiferroic VHfO\texorpdfstring{$_4$}{4} with Strain-Tunable Magnetism}

\author{Qisheng Yu}
\thanks{These three authors contributed equally.}
\affiliation{Zhejiang University, Hangzhou 310058, China}
\affiliation{Department of Physics, School of Science, Westlake University, Hangzhou 310030, China}
\author{Tianyuan Zhu}
\thanks{These three authors contributed equally.}
\affiliation{Department of Physics, School of Science, Westlake University, Hangzhou 310030, China}
\affiliation{Institute of Natural Sciences, Westlake Institute for Advanced Study, Hangzhou 310024, China}
\author{Boyu Liu}
\thanks{These three authors contributed equally.}
\affiliation{Key Laboratory of Computational Physical Sciences (Ministry of Education), Institute of Computational Physical Sciences, State Key Laboratory of Surface Physics and Department of Physics, Fudan University, Shanghai 200433, P. R. China}

\author{Hongjun Xiang}
\affiliation{Key Laboratory of Computational Physical Sciences (Ministry of Education), Institute of Computational Physical Sciences, State Key Laboratory of Surface Physics and Department of Physics, Fudan University, Shanghai 200433, P. R. China}

\author{Shi Liu}
\email{liushi@westlake.edu.cn}
\affiliation{Department of Physics, School of Science, Westlake University, Hangzhou 310030, China}
\affiliation{Institute of Natural Sciences, Westlake Institute for Advanced Study, Hangzhou 310024, China}

\begin{abstract}
The coexistence of ferroelectricity and magnetism in a single-phase oxide is rare because the electronic requirements for these two orders are often incompatible. Here, using first-principles calculations and parallel-tempering Monte Carlo simulations, we propose stoichiometric VHfO$_4$ as a hafnia-derived multiferroic that overcomes this constraint through ordered cation design rather than dilute magnetic doping.
We found that VHfO$_4$ \note{adopts a symmetry-lowered polar \Pc{} structure derived from the ferroelectric \Pca{} phase}, with layered V/Hf ordering, \note{local} dynamical stability, and switchable ferroelectricity with a large spontaneous polarization.
The ordered V sublattice introduces competing exchange interactions that favor an antiferromagnetic ground state at zero strain. Epitaxial strain further drives transitions into additional phases, including a noncollinear spiral-like state and a predominantly in-plane antiferromagnetic state. We also find that out-of-plane lattice distortions along the polar axis strongly modify the exchange interactions and magnetic phase stability, indicating a strain-mediated pathway for electric-field control of magnetism.
These results establish VHfO$_4$ as a promising \note{artificial-superlattice candidate} for exploring multiferroicity and magnetoelectric coupling in hafnia-based oxides.
\end{abstract}

\maketitle

\clearpage

\section{Introduction}

Multiferroic materials, which combine more than one ferroic order such as ferroelectricity and ferromagnetism, offer exciting possibilities for future electronic devices. Integrating these materials into silicon-compatible platforms could enable next-generation non-volatile logic and memory technologies, where electric fields control magnetic states~\cite{Schmid94p317,Spaldin19p203,Manipatruni19p35,Ramesh07p21}. However, realizing strong ferroelectric and magnetic properties in a single material is a major challenge. This difficulty arises because the electronic structures that support each property tend to conflict~\cite{Hill00p6694}. Magnetism typically requires  partially filled $d$-shells to host local magnetic moments~\cite{Goodenough63p394,Coey10p}, whereas  conventional ferroelectricity prefers empty $d^0$ configurations to enable strong cation–anion $p$--$d$ hybridization that stabilizes polar off-centering~\cite{Cohen92p136}. To overcome this issue, various mechanisms have been developed in complex perovskite oxides. For example, BiFeO$_3$ is a widely studied multiferroic where ferroelectricity originates from the stereochemically active $6s^2$ lone pair on Bi$^{3+}$~\cite{Wang03p1719}, which induces a non-centrosymmetric distortion and ferroelectric polarization. Simultaneously, the Fe$^{3+}$ ions (with a half-filled $3d^5$ configuration) exhibit G-type antiferromagnetism.
In contrast, TbMnO$_3$ represents a type-II multiferroic, where ferroelectricity emerges from a non-collinear spiral spin structure of Mn$^{3+}$ ions at low temperatures~\cite{Kimura03p55}. This magnetic ordering breaks inversion symmetry and induces polarization via the inverse Dzyaloshinskii–Moriya interaction. \note{Related theoretical and review works have also emphasized the broader role of spin-orbit and spin-lattice couplings in multiferroics and magnetoelectric phenomena~\cite{Liu12p373,Zhang12p408,Dong19p629}.}
Recently, a type-III multiferroelectric was proposed in a single-atom adsorbed monolayer Cu-InI$_3$~\cite{Jiang25p196801}. In this system, the ferroelectric lattice distortion reduces interatomic orbital overlap and narrows the electronic bands. As a result, the density of states at the Fermi level is markedly enhanced, inducing a Stoner instability and driving a transition from a nonmagnetic to a ferromagnetic state dominated by In-$s$ orbitals. This ``ferroelectricity-driven magnetism" represents a distinct pathway, enabling efficient electric control of magnetic order without relying on transition-metal $d$-electrons.

Despite these advances, integrating such complex oxides with standard complementary metal-oxide-semiconductor (CMOS) technology remains a significant challenge. These materials often require high-temperature synthesis and epitaxial growth on lattice-matched substrates, which are incompatible with conventional silicon processing. Furthermore, scaling down multiferroic oxides to nanometer dimensions could degrade their ferroic properties due to increased surface and interface effects. Issues such as chemical interdiffusion and poor interface quality further hinder reliable device performance, limiting their practical application in miniaturized and densely integrated circuits.

Recent progress in fluorite-structured oxides, particularly hafnia-based systems~\cite{Boscke11p102903,Lee20p1343,Luo20p1391,Kim21peabe1341}, provides a unique opportunity to develop CMOS-compatible multiferroics. Unlike conventional perovskite ferroelectrics, the ferroelectricity in HfO$_2$ arises from a non-centrosymmetric distortion of oxygen sublattices in the polar orthorhombic phase (space group \Pca{}), rather than relying on $d^0$ transition metal cations. This distinct mechanism relaxes the conventional $d^0$ rule constraint, suggesting that ferroelectricity and magnetism could, in principle, coexist within the same material system.
However, the polar orthorhombic phase of HfO$_2$ is thermodynamically metastable relative to the monoclinic ground state~\cite{Huan14p064111} and requires stabilization through extrinsic factors such as dopant incorporation~\cite{Boscke11p102903,Mller11p114113}, surface energy effects~\cite{Materlik15p134109}, or strain engineering~\cite{Shiraishi16p262904,Batra17p4139,Wei18p1095,Fina21p1530}. In this context, we propose that magnetic cation doping may play a dual role: promoting the stabilization of the polar phase and simultaneously introducing magnetic functionality. This approach could enable the design of single-phase multiferroics that are fully compatible with CMOS technology.

Previous efforts to introduce magnetism into hafnia have largely relied on dilute magnetic doping, such as V-doped HfO$_2$~\note{\cite{Turquat99p3,Ansari25p2702}}.
While experimental studies have reported magnetic signatures, the low concentration and random distribution of dopants typically lead to weak exchange interactions~\cite{Coey05p173,Abraham05p252502,Tomida06p142902,Zhou15p240}. To overcome these limitations, a promising direction involves moving beyond dilute doping toward the design of stoichiometric magnetic compounds or engineered superlattice structures, where magnetic cations are periodically ordered. These systems can support stronger, long-range exchange coupling by establishing well-defined magnetic sublattices, thereby enhancing the potential for robust magnetoelectric effects within a CMOS-compatible platform. \note{Related Hf--V--O oxides and hafnia-based artificial-layer structures have been explored experimentally~\cite{Fleer19p21354,Ravensburg20p172,Lehninger23p2200108,Li2025p6417}, motivating ordered hafnia-derived architectures even though a perfectly layered VHfO$_4$ superlattice has not yet been synthesized.}

In this work, we employ first-principles calculations to investigate the structural, electronic, and magnetic properties of the stoichiometric \note{ordered} compound VHfO$_4$, corresponding to \note{a layered (HfO$_2$)$_{0.5}$/(VO$_2$)$_{0.5}$ superlattice}.
\note{The 50\% V content is therefore part of the ordered-compound design rather than a variable dilute-dopant concentration.}
Our results show that VHfO$_4$ \note{relaxes into a polar \Pc{} phase derived from the \Pca{} hafnia structure}, preserving the non-centrosymmetric lattice distortions and ferroelectric switching pathways of pure HfO$_2$, while incorporating a dense network of V$^{4+}$ ($d^1$) magnetic ions.
The periodic arrangement of magnetic cations enables the emergence of long-range magnetic order. We find that the low-symmetry polar lattice gives rise to competing nearest- and next-nearest-neighbor interactions, resulting in magnetic frustration that stabilizes non-collinear spin textures, including spiral configurations. Additionally, these magnetic states exhibit strong sensitivity to epitaxial strain, which modulates the exchange couplings and drives transitions between collinear and non-collinear magnetic phases. These findings demonstrate that ordered hafnia-based compounds such as VHfO$_4$ offer a promising platform for realizing strain-tunable multiferroicity and electric-field control of magnetic order via magnetoelastic coupling.

\section{Computational Methods}

First-principles calculations were performed using the plane-wave projector augmented wave (PAW) method~\cite{Blochl94p17953} as implemented in the Vienna Ab-initio Simulation Package (VASP)~\cite{Kresse96p11169, Kresse96p15}. The exchange-correlation interaction was treated within the Perdew-Burke-Ernzerhof (PBE) generalized gradient approximation~\cite{Perdew96p3865}. To account for the strong on-site Coulomb repulsion associated with the localized V-$3d$ electrons, the DFT$+U$ formalism was employed with an effective Hubbard parameter of $U_{\rm eff} = 5.0$ eV~\cite{Anisimov91p943, Dudarev98p1505}. \note{Hybrid-functional HSE06 calculations were used as an electronic-structure benchmark for the band gap and band dispersion. Spin-orbit-coupled calculations were further carried out in a 96-atom $2\times2\times2$ supercell to estimate the magnetocrystalline anisotropy energy (MAE).}
Structural optimizations and property calculations were performed using a 12-atom unit cell representing stoichiometric VHfO$_4$.
A plane-wave energy cutoff of 800 eV was used, and the Brillouin zone was sampled with a $\Gamma$-centered Monkhorst–Pack 4$\times$4$\times$4 $k$-point mesh of appropriate density to ensure convergence. All atomic positions and lattice parameters were fully relaxed until the atomic forces converged below the threshold of 0.001 eV/\AA.
To assess the dynamical stability of the predicted structure, phonon dispersion relations were calculated using the finite displacement method as implemented in the \texttt{PHONOPY} package~\cite{Togo15p1}. \note{Additional phonon calculations were performed for representative strained AFM-i and NC structures using $2\times2\times2$ supercells.} Ferroelectric switching pathways and associated energy barriers were evaluated using the Variable-Cell Nudged Elastic Band (VCNEB) method~\cite{Qian13p2111}, as implemented in \texttt{USPEX}~\cite{Oganov06p244704,Lyakhov13p1172,Oganov11p227}, which enables simultaneous relaxation of both lattice vectors and internal atomic coordinates along the minimum energy path. 

To investigate the magnetic phase diagram, we mapped the DFT total energies onto a classical Heisenberg model: $H = -\sum_{i<j} J_{ij} \mathbf{S}_i \cdot \mathbf{S}_j$, where $J_{ij}$ denotes the exchange coupling constant between magnetic moments located at sites $i$ and $j$, and $\mathbf{S}_i$ represents the classical spin vector at site $i$. The exchange coupling constants $J_{ij}$ were extracted using the four-state method~\cite{Xiang11p224429,Xiang12p823}, which involves computing the total energies of four distinct spin configurations for a given pair of interacting spins. For a specific pair $(i, j)$, the coupling is derived as:
\begin{equation}
J_{ij} = \frac{E_{ij,\uparrow\uparrow} + E_{ij,\downarrow\downarrow} - E_{ij,\uparrow\downarrow} - E_{ij,\downarrow\uparrow}}{4}
\nonumber
\end{equation}
where $E_{ij,\sigma_i\sigma_j}$ is the total energy of the system with spins $\sigma_i$ and $\sigma_j$ aligned in the specified directions, while all other spins remain fixed. This approach assumes a collinear spin framework and treats the magnetic moments as classical vectors.

Magnetic ground states and transition temperatures were evaluated using parallel tempering Monte Carlo (PTMC) simulations, implemented via the \texttt{PASP} package~\cite{Lou21p154}, which allows for efficient sampling of complex spin configurations in frustrated magnetic systems. The simulations were initialized with exchange coupling constants $J_{ij}$ obtained from DFT calculations and were carried out over a temperature range of 1 K to 50 K. A total of 800 exchange steps were used in each PTMC run, during which configurations between adjacent temperature replicas were periodically exchanged to enhance ergodicity and avoid trapping in local energy minima~\cite{Hukushima96p1604,Earl05p3910,Hansmann97p140}. The first 400 steps were designated for thermal equilibration to ensure convergence toward statistically representative configurations. Subsequently, 200 MC block steps were used for statistical sampling of spin configurations and thermodynamic observables, such as spin structure factors and specific heat, from which the magnetic ordering temperature and dominant spin textures were extracted. 

\section{Results and Discussion}
\subsection{Thermodynamic Stability}
\begin{figure}[t]
\centering
\includegraphics[width=0.9\textwidth]{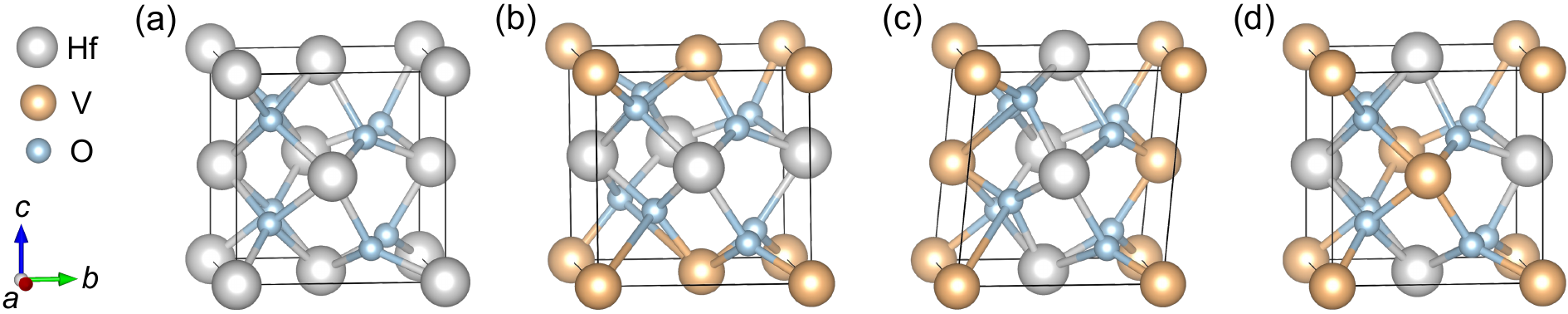}
\caption{(a) Crystal structure of the parent ferroelectric \Pca{} phase of HfO$_2$. (b)--(d) Configurations of stoichiometric VHfO$_4$ exhibiting different cation ordering patterns within the unit cell.}
\label{str-doping}
\end{figure}

Ferroelectricity in hafnia-based oxides is widely attributed to the orthorhombic \Pca{} phase (Fig.~\ref{str-doping}a), which features locally displaced oxygen atoms. Importantly, the \Pca{} phase is metastable, with energy higher than the monoclinic $P2_1/c$ and antiferroelectric-like $Pbca$ phases. 
To investigate how vanadium incorporation affects structural stability, we begin by examining the influence of V/Hf cation ordering within the \Pca{} framework. For the 12-atom unit cell considered, three symmetry-inequivalent cation configurations are possible, as depicted in Fig.~\ref{str-doping}b–d. Among these, a layered ordering of Hf and V ions along the polar $c$ axis (Fig.~\ref{str-doping}b) emerges as the lowest-energy configuration. In contrast, the alternative arrangements shown in Fig.~\ref{str-doping}c and Fig.~\ref{str-doping}d are either energetically unfavorable or dynamically unstable, relaxing to non-polar structures upon optimization. The most stable configuration can be interpreted as a (HfO$_2$)$_{0.5}$/(VO$_2$)$_{0.5}$ superlattice, which maintains the essential polar distortions characteristic of the \Pca{} phase. \note{We further verified that this polar-axis layered ordering remains lowest in energy under the representative AFM-i and NC strain states discussed below (Table~S1).}

\begin{figure}[t]
\centering
\includegraphics[width=0.65\textwidth]{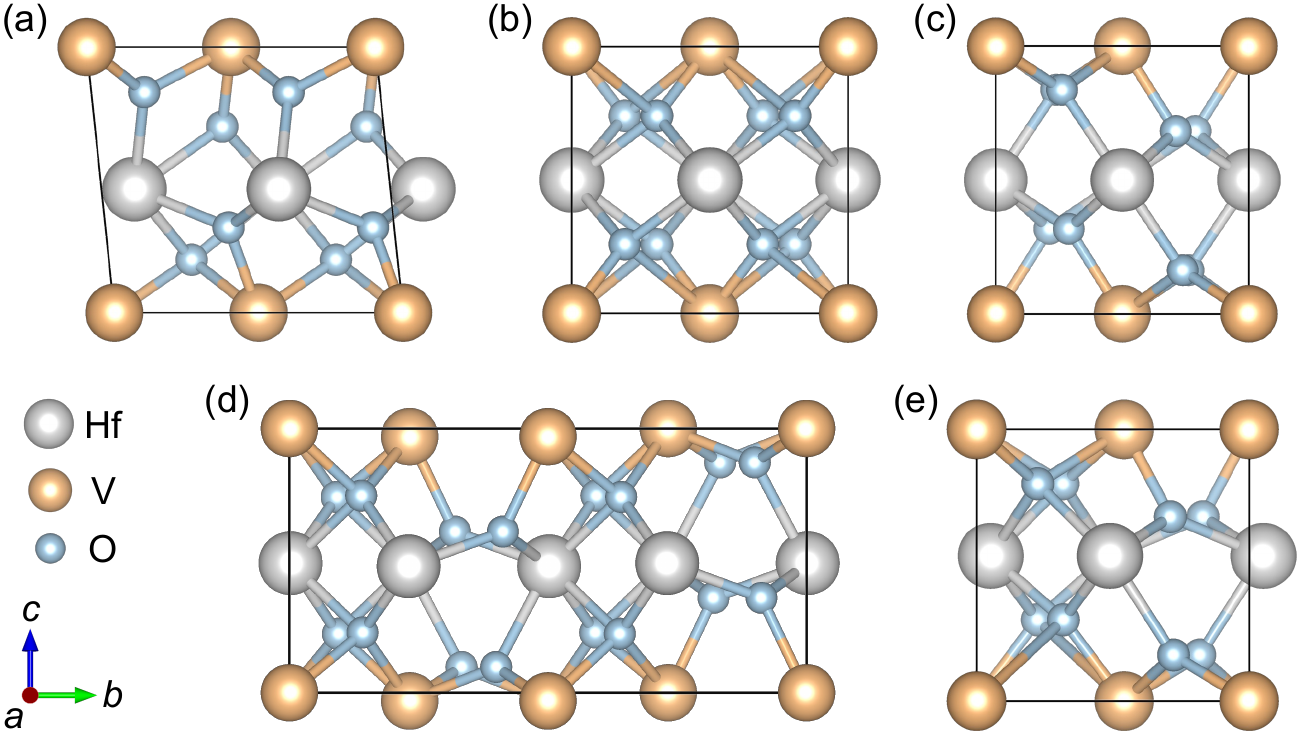}
\caption{Crystal structures of the layered VHfO$_4$ system. (a)--(e) Schematic crystal structures of VHfO$_4$ obtained through structural optimization starting from five distinct HfO$_2$ polymorphic templates: (a) $P2_1/c$, (b) $P4_2/nmc$, (c) $Pbcn$, (d) $Pbca$, and (e) \Pca{}.}
\label{str-phase}
\end{figure}

Having identified this layered cation ordering as the most favorable within the \Pca{} phase, we adopt it as a representative V distribution to evaluate the relative stability of VHfO$_4$ polymorphs derived from common HfO$_2$ structures. As shown in Fig.~2, these include the monoclinic $P2_1/c$, tetragonal $P4_2/nmc$, and orthorhombic \Pca{} and antiferroelectric-like $Pbcn$ and $Pbca$ phases (see Fig.~\ref{str-phase})~\cite{Huan14p064111,Sang15p162905,Valmalette98p168,Caravaca05p5795}. For simplicity, we refer to the VHfO$_4$ structures by the names of their parent HfO$_2$ polymorphs, even though V substitution may alter the symmetry.
For each polymorph, we compared the energies of three magnetic states: nonmagnetic (NM), ferromagnetic (FM), and antiferromagnetic (AFM). Table 1 presents the lattice parameters for these polymorphs across distinct magnetic states, along with the energy difference ($\Delta E$) relative to the nonmagnetic $P2_1/c$ phase. Although magnetism has a negligible effect on the structural parameters, stability depends strongly on the magnetic state; generally, the FM state is the most stable for a 12-atom unit cell.
Notably, the structure initialized from the polar \Pca{} phase relaxes into a monoclinic \note{polar} unit cell \note{with space group \Pc{} (No. 7)}, while preserving atomic displacements similar to the original \Pca{} configuration. We therefore \note{use ``\Pca-like'' to denote a cation-ordered, symmetry-lowered \Pc{} derivative of the polar \Pca{} hafnia phase~\cite{Cernov26parXiv}.}
Furthermore, we find that the \Pca-like phase of VHfO$_4$ is more stable than both the $P2_1/c$ and $Pbca$ phases \note{within the considered ordered structural models.}
This contrasts with HfO$_2$, where $P2_1/c$ is the ground state, followed by $Pbca$. These results suggest that the inclusion of V layers stabilizes the polar \Pca-like phase while simultaneously inducing magnetism.

\begin{figure}[t]
\centering
\includegraphics[width=0.75\textwidth]{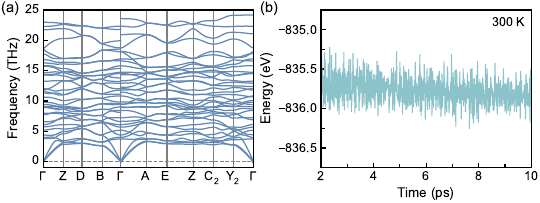}
\caption{Dynamical and thermal stability of the \Pca-like VHfO$_4$ phase. (a) Phonon dispersion spectrum, indicating the absence of imaginary modes and confirming dynamical stability. (b) Evolution of the total energy during $ab$ $initio$ molecular dynamics simulations at 300 K. The data are plotted for the time interval of 2--10 ps, where the stable fluctuation of energy around the equilibrium value confirms the thermal stability of the structure at room temperature.}
\label{phononband}
\end{figure}

We assess the dynamical stability of the FM \Pca-like VHfO$_4$ structure using a two-step approach. First, we calculate the phonon dispersion spectrum to test stability within the harmonic approximation. For a structure to be dynamically stable, the potential energy surface must be a local minimum, meaning the lattice must exhibit restoring forces against any small perturbation. This condition requires the absence of imaginary modes in the phonon spectrum. As shown in Fig.~\ref{phononband}a, our calculations reveal that \Pca-like VHfO$_4$ is free from imaginary frequencies across the entire Brillouin zone, confirming its intrinsic stability against small displacements.

Because phonon calculations represent zero-temperature conditions, we further employ $ab$ $initio$ molecular dynamics (AIMD) simulations to evaluate the structural integrity under finite-temperature thermal fluctuations and anharmonic effects. We simulated a $2\times2\times2$ supercell (96 atoms) at 300~K. As plotted in Fig.~\ref{phononband}b, the total energy oscillates around a constant equilibrium value without drifting, and the atomic structure remains intact throughout the simulation time.
These results indicate that the \Pca-like VHfO$_4$ structure, characterized by a layered ordering of Hf and V ions along the polar $c$ axis, is \note{locally and dynamically stable} and is a promising \note{artificial-superlattice candidate} for multiferroicity. \note{This local stability should be distinguished from full bulk thermodynamic phase stability over a chemical-potential window. Experimentally, V incorporation in hafnia, Hf--V--O ternary oxides, and hafnia-based oxide superlattices have all been reported~\cite{Turquat99p3,Ansari25p2702,Fleer19p21354,Ravensburg20p172,Lehninger23p2200108,Li2025p6417}, providing motivation for future exploration of ordered VHfO$_4$-type architectures.}

\subsection{Electronic Structure}
\begin{figure}[t]
\centering
\includegraphics[width=1.0\textwidth]{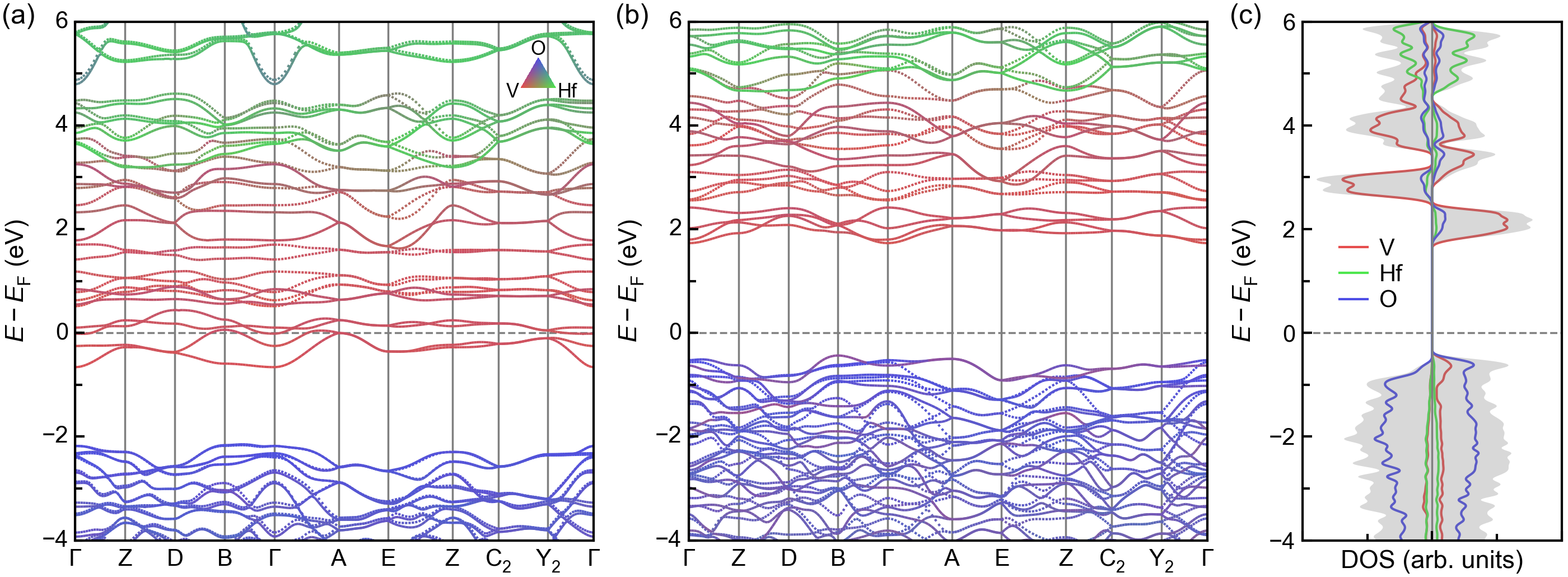}
\caption{Impact of the Hubbard $U$ correction on the electronic structure of VHfO$_4$. (a) Spin-polarized band structure calculated using standard DFT (without $U$), exhibiting a metallic character. (b) Band structure derived from DFT+$U$ calculations, where the inclusion of on-site Coulomb interactions opens a significant band gap of approximately 2.17 eV. (c) Projected density of states (PDOS) corresponding to the DFT+$U$ result, resolving the elemental contributions near the band edges.}
\label{band}
\end{figure}

We find that the electronic structure of FM \Pca-like VHfO$_4$ displays characteristics of a Mott insulating state. Figure~\ref{band} compares the spin-polarized electronic band structures computed using DFT+$U$ for two representative values: $U_{\rm eff} = 0$~eV and $U_{\rm eff} = 5.0$~eV. In the absence of on-site Coulomb interactions ($U_{\rm eff} = 0$~eV), the system is metallic (Fig.~\ref{band}a), consistent with the nominal V$^{4+}$ ($3d^1$) electronic configuration. Within a simple band picture, the single V-$3d$ electron partially occupies the conduction band, yielding a metallic state.
However, the inclusion of a finite on-site Hubbard interaction ($U_{\rm eff} = 5.0$~eV) leads to the opening of a gap of approximately 2.17~eV (Fig.~\ref{band}b), thereby driving a transition from a metallic to an insulating state.
Hybrid-functional HSE06 calculations predict a finite band gap of 2.48~eV for FM \Pca-like VHfO$_4$\note{, and the PBE+$U$ band dispersions near the VBM and CBM remain consistent with the HSE06 results (Fig.~S1). Together with the $U_{\rm eff}$-dependent lattice parameters and band gaps listed in Table~S2, this comparison supports the use of $U_{\rm eff} = 5.0$~eV in the present DFT+$U$ calculations.}

This insulating behavior is characteristic of a Mott-Hubbard insulator, where strong on-site electron correlations overwhelm the kinetic energy gain associated with band dispersion~\cite{Mott49p416,Zaanen85p418}. A detailed analysis of the PDOS reveals that both the valence band maximum (VBM) and the conduction band minimum (CBM) consist mainly of V-$3d$ and O-$2p$ states. The band gap arises from the splitting of the V $3d$ manifold into occupied lower Hubbard bands and unoccupied upper Hubbard bands. 
The localization of the V $3d$ electron, driven by strong on-site Coulomb repulsion, highlights the Mott-Hubbard nature of the insulating state in FM \Pca-like VHfO$_4$.

\subsection{Ferroelectricity}
To assess whether the polar structure of VHfO$_4$ supports switchable ferroelectricity, we compute the minimum energy paths (MEPs) for polarization reversal using the variable-cell nudged elastic band (VCNEB) method~\cite{Huang22p144106,Yu23p142902}. Following established approaches for pure HfO$_2$, we consider three representative switching pathways: two ``shift-inside'' (SI) modes, involving oxygen displacements between adjacent cation sub-layers, and one ``shift-across'' (SA) mode, in which oxygen ions traverse the cation plane~\cite{Wei22p154101,Choe21p8}.

\begin{figure}[t]
\centering
\includegraphics[width=1.0\textwidth]{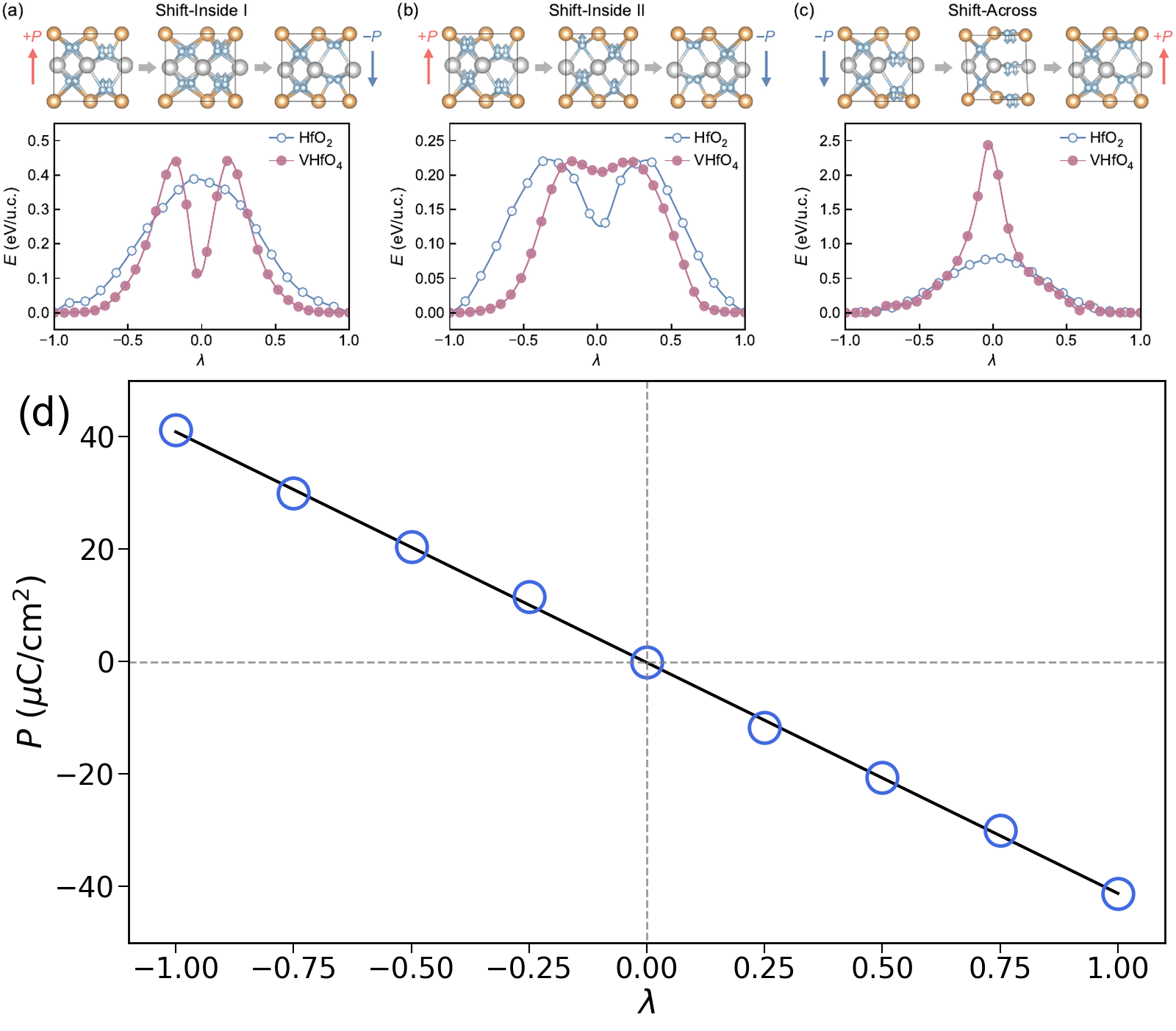}
\caption{Comparison of minimum energy pathways (MEPs), atomic switching mechanisms, and \note{Berry-phase polarization evolution} for polarization reversal in \Pca-like VHfO$_4$ and pure HfO$_2$. (a)--(c) Calculated energy profiles using the variable-cell nudged elastic band (VCNEB) method for three distinct pathways: shift-inside I (SI-I), shift-inside II (SI-II), and shift-across (SA). The schematics at the top illustrate the corresponding atomic motions, \note{and the arrows indicate the dynamically assigned polarization directions.} \note{(d) Polarization evolution along the SI-II pathway in \(\mu\mathrm{C}/\mathrm{cm}^2\), showing the continuous change between the two opposite polar directions.}}
\label{NEB}
\end{figure}

The SA pathway (Fig.~\ref{NEB}c) exhibits a prohibitively high energy barrier of approximately 2.4~eV, likely due to the significant energetic cost of breaking short V--O bonds. In contrast, the \note{SI-II} pathway (Fig.~\ref{NEB}b), in which both three-fold and four-fold coordinated oxygen ions migrate, displays a much lower barrier of 0.25~eV, indicating it is kinetically accessible. Notably, this barrier is comparable to those reported for undoped HfO$_2$ ($\approx$0.22~eV)~\cite{Wei22p154101} and conventional ferroelectrics such as PbTiO$_3$ ($\approx$0.20~eV)~\cite{Cohen92p136}. The \note{SI-I} pathway (Fig.~\ref{NEB}a) in VHfO$_4$ features a transient state that is lower in energy compared to the corresponding configuration in HfO$_2$, although the overall barrier remains similar.
Additionally, based on Berry-phase calculations \note{along the SI-II pathway, the polarization evolves continuously from positive to negative values, indicating reversal of the polarization direction} \note{(Fig.~\ref{NEB}d). The polarization is reported in \(\mu\mathrm{C}/\mathrm{cm}^2\), consistent with Fig.~\ref{NEB}d. In the modern theory of polarization~\cite{Ghosez22p325}, the meaningful quantity is the change in polarization along a continuous switching path, and the arrows in Fig.~\ref{NEB}a--c are therefore assigned according to this dynamical polarization evolution.}
These results suggest that the incorporation of a magnetic V sublattice does not hinder polarization switching. VHfO\(_4\) preserves the robust, switchable polar character of the fluorite host structure via the energetically favorable \note{SI-II} mechanism, establishing it as a promising ferroelectric.

\subsection{Magnetism}
\begin{figure}[t]
\centering
\includegraphics[width=1.0\textwidth]{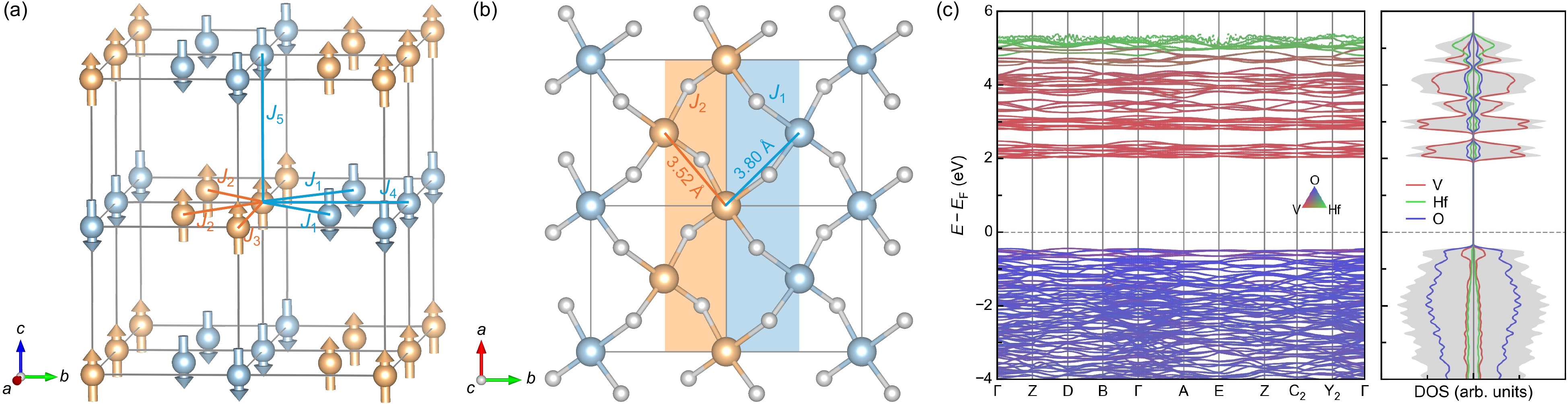}
\caption{Magnetic ground state and exchange interactions of the V sublattice in layered VHfO$_4$. The schematic illustrates the calculated magnetic ground state of the system. The schematic shows the calculated magnetic ground-state configuration of the system, with only the V atoms displayed. The spheres represent V atoms, and the arrows indicate the orientations of their local magnetic moments. The network of magnetic exchange interactions is labeled as $J_1$--$J_5$, with values of $J_1=-2.0$, $J_2=13.73$, $J_3=1.04$, $J_4=-0.17$, and $J_5=-0.21$(meV). The background colors distinguish the nature of the magnetic coupling: the orange background represents ferromagnetic (FM) interactions, while the blue background represent antiferromagnetic (AFM) interactions.}
\label{AFMorder}
\end{figure}

Determining the magnetic ground state of VHfO$_4$ is nontrivial because the system crystallizes in a low-symmetry \Pca-like structure. As summarized in Table~\ref{str}, the in-plane lattice is anisotropic (\(a>b\)), which lowers the symmetry and gives rise to five inequivalent exchange pathways of comparable length scale (Fig.~\ref{AFMorder}a). This complexity makes the magnetic energetics sensitive to the size of the simulation cell and to the range of spin configurations included.
The results reported in Sec.~III.A were obtained from DFT calculations in the primitive 12-atom cell and favored a ferromagnetic (FM) state. That result, however, is not conclusive, because the small magnetic cell restricts the set of ordering wave vectors allowed by periodic boundary conditions. In particular, the primitive cell can only accommodate \(q=0\) spin arrangements and therefore excludes antiferromagnetic (AFM) states with \(q\neq 0\) by construction.

To access these more general magnetic configurations, we enlarged the magnetic cell to a \(2\times2\times2\) supercell and extracted the exchange constants \(J_{ij}\) using the four-state method. The structural anisotropy (\(a>b\)) splits the in-plane nearest-neighbor V--V interactions into two distinct channels: a weak antiferromagnetic coupling, \(J_1 \approx -2\)~meV, and a much stronger ferromagnetic coupling, \(J_2 \approx 13.7\)~meV. The remaining interactions are substantially smaller, with next-nearest-neighbor exchanges \(J_3 \approx 1.0\)~meV and \(J_4 \approx -0.2\)~meV, and a weak antiferromagnetic interlayer coupling along the \(c\) axis, \(J_5 \approx -0.2\)~meV. \note{A representative $U_{\rm eff}$-sensitivity test for AFM, NC, and AFM-i states shows that changing $U_{\rm eff}$ from 4.0 to 6.0~eV only moderately changes the numerical values of the exchange constants and does not alter the dominant role of $J_2$ (Table~S3).}

Using these exchange parameters, parallel-tempering Monte Carlo simulations using $6\times6\times2$ converge to the ground state shown in Fig.~\ref{AFMorder}a. The resulting magnetic order can be understood most naturally in terms of the hierarchy between \(J_2\) and \(J_1\). Within the \(ab\) plane, the shorter V--V links associated with \(J_2\) form zigzag chains, and the strong ferromagnetic interaction \(J_2>0\) aligns spins parallel along each chain. Neighboring chains are then coupled antiferromagnetically through the weaker interaction \(J_1<0\), producing an overall collinear AFM pattern with local moments oriented along the \(c\) axis. As shown in Fig.~\ref{AFMorder}b, this interpretation is consistent with the bond geometry: the V--V distance for \(J_2\) is 3.52~\AA, notably shorter than the 3.80~\AA\ separation corresponding to \(J_1\). Along the \(c\) axis, adjacent layers also remain antiferromagnetically aligned, consistent with \(J_5<0\).
This AFM state is lower in energy than the FM state by 0.09~meV/atom. We further calculated its electronic structure and found an insulating band gap of 2.46~eV (Fig.~\ref{AFMorder}c). The atom-resolved density of states likewise shows features consistent with the AFM ground state (Fig.~\ref{AFMorder}d). \note{Spin-orbit-coupled calculations in 96-atom \(2\times2\times2\) supercells give maximum MAE values of 1.34, 0.71, and 0.76~meV/cell for the AFM-i, NC, and AFM states, respectively, corresponding to only 0.007--0.014~meV/atom (Table~S4). These values are much smaller than the leading exchange scale, especially \(J_2\), indicating that the magnetic phase stability is governed mainly by exchange interactions. The absence of inversion symmetry still allows Dzyaloshinskii-Moriya interactions by symmetry~\cite{Yang23p43}, but the MAE scale suggests that spin-orbit effects provide a secondary anisotropic correction rather than the primary driving force for the calculated magnetic phases.}

\subsection{Epitaxial Strain Effect}
\begin{figure}[t]
\centering
\includegraphics[width=1.0\textwidth]{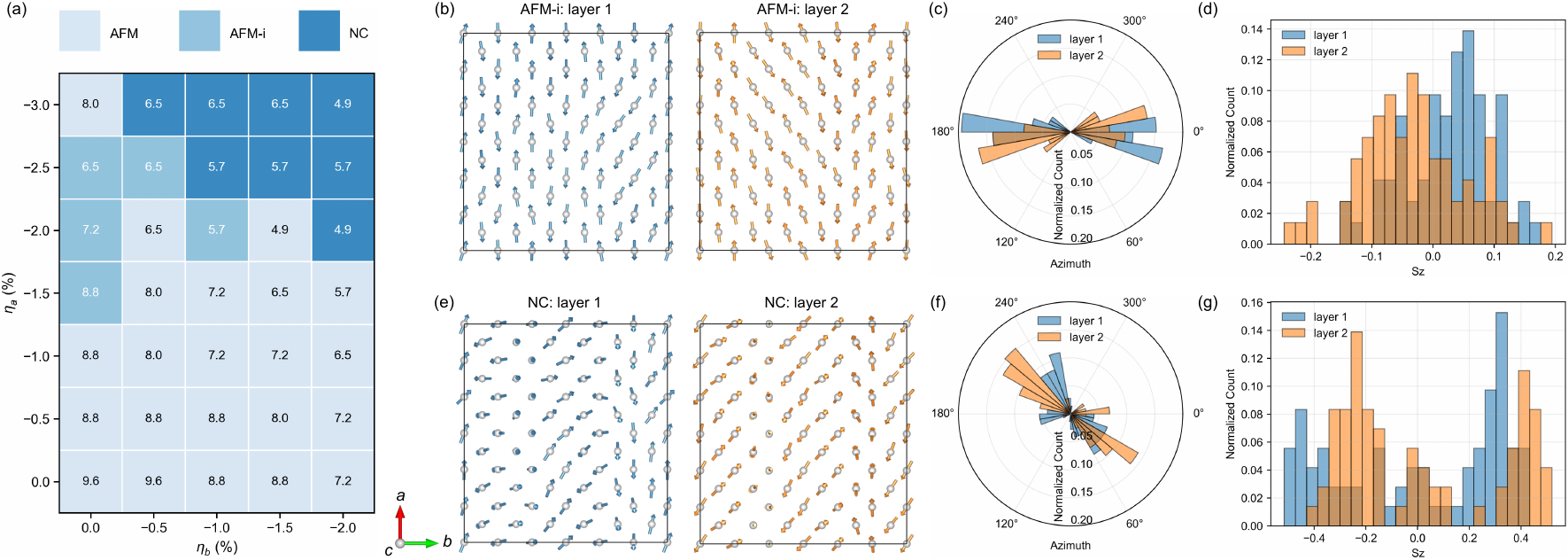}
\caption{(a) Magnetic phase diagram of $Pca2_1$-like VHfO$_4$ as a function of in-plane strains $\eta_a$ and $\eta_b$. The numeric values within each grid indicate the calculated magnetic transition temperatures ($T_{\text{trans}}$) in Kelvin. (b) Schematic of the AFM-i configuration stabilized at $\eta_a = -2.5\%$ and $\eta_b = -0.5\%$, with the azimuthal-angle distribution within the $ab$ plane and the distribution of the out-of-plane spin component ($S_z$) shown in (c) and (d). (e) Schematic of the NC configuration at $\eta_a = -2.5\%$ and $\eta_b = -2.0\%$, with azimuthal-angle and $S_z$ distributions shown in (f) and (g).}
\label{mag}
\end{figure}

The pronounced coupling between lattice distortions and magnetic exchange interactions suggests that epitaxial strain may provide an effective route to control the magnetic ground state. To test this idea, we mapped the strain-dependent magnetic phase space using parallel-tempering Monte Carlo simulations parameterized by DFT-derived Heisenberg exchange constants \(J_{ij}\). The resulting phase diagram, shown in Fig.~\ref{mag}a, reveals a clear strain-driven evolution of magnetic order and demonstrates that relatively simple collinear order in the unstrained system can be transformed into more complex spin textures under biaxial strain.
\note{Such strain engineering is experimentally relevant for hafnia-based films, where substrate constraints, interfaces, and artificial layering can stabilize polar phases and tune their functional response~\cite{Wei18p1095,Fina21p1530,Lehninger23p2200108,Li2025p6417}. The present strain calculations therefore provide a practical route for assessing how a layered VHfO$_4$ architecture could be manipulated in thin-film or superlattice form.}

At zero strain, the system adopts the collinear AFM ground state shown in Fig.~\ref{AFMorder}. Under compressive strain in the \(ab\) plane, however, the magnetic ground state changes qualitatively, stabilizing two additional phases: a noncollinear spiral-like state (denoted NC) and an in-plane antiferromagnetic state (denoted AFM-i). Figures~\ref{mag}b and \ref{mag}e display these two configurations in a \(6\times6\times2\) supercell, with both layers along the \(c\) axis shown explicitly.
\note{To verify that these representative strained structures remain locally stable, we calculated phonon spectra for AFM-i at \((\eta_a,\eta_b)=(-2.5\%,-0.5\%)\) and NC at \((\eta_a,\eta_b)=(-2.5\%,-2.0\%)\) using \(2\times2\times2\) supercells. No imaginary phonon modes are found in either case (Fig.~S2), confirming dynamical stability for the strain points used to represent the two additional magnetic phases.}

In the AFM-i phase, the spin pattern is best described as predominantly antiferromagnetic with moments lying mainly in the \(ab\) plane, pointing largely along the \(b\) axis. Although the plotted Monte Carlo configuration is not perfectly collinear, the small canting is most likely a consequence of the stochastic nature and finite convergence tolerance of the Monte Carlo sampling,. The robust feature of AFM-i is therefore the underlying antiferromagnetic arrangement together with the reorientation of the spin moments from the \(c\) axis into the \(ab\) plane.
In the NC phase, the zigzag V--V chains remain antiferromagnetically coupled between adjacent layers along \(c\), while the in-plane spin arrangement becomes noncollinear. Specifically, spins on neighboring chains are no longer strictly parallel or antiparallel, but instead vary gradually in space, giving rise to a spiral-like modulation. 

The emergence of these strain-induced phases can be understood as a consequence of enhanced magnetic frustration. Compressive strain modifies the V--O--V bond lengths and bond angles, which in turn changes the balance among competing exchange interactions, particularly \(J_1\) and \(J_2\). As these couplings become closer in magnitude while favoring different spin alignments, the system can lower its energy by developing either a genuinely noncollinear state, as in NC, or an in-plane AFM state with weak residual canting, as in AFM-i. The calculated ordering temperatures of these strain-stabilized phases lie in the range of approximately 5--20~K (Fig.~\ref{mag}a). Although these transition temperatures remain relatively low, the results establish the key principle that epitaxial strain can be used to engineer and switch magnetic order in this fluorite oxide.

To characterize the spin textures in the two strain-stabilized phases more quantitatively, we further analyzed the distribution of spin orientations in AFM-i and NC configurations by computing the azimuthal-angle distribution within the plane and the distribution of the out-of-plane spin component \(S_z\).
The AFM-i state shows a much narrower azimuthal distribution, with dominant weight around two nearly opposite directions (Fig.~\ref{mag}c), as expected for a predominantly antiferromagnetic arrangement with an approximately collinear in-plane axis. Its \(S_z\) distribution is mostly concentrated near \(S_z=0\) (Fig.~\ref{mag}d), demonstrating that the spins lie mainly in the \(ab\) plane .
For the NC state, the azimuthal distribution is broadly spread over multiple angles rather than concentrated at only one or two symmetry-related directions (Fig.~\ref{mag}f). This behavior is consistent with a continuously rotating in-plane spin texture and therefore supports the spiral-like character inferred from the real-space spin pattern. At the same time, the \(S_z\) distribution exhibits substantial weight over a relatively wide range of positive and negative values (Fig.~\ref{mag}g), indicating that the spins retain appreciable out-of-plane components and are not confined to a single plane.

\subsection{Potential Magnetoelectric Coupling}
Finally, we investigated the coupling between ferroelectricity and magnetism to evaluate whether the magnetic order in VHfO$_4$ can be controlled electrically through the converse piezoelectric effect. Since the spontaneous polarization is oriented along the \(c\) axis, an external electric field applied in this direction is expected to modify both the polarization and the associated out-of-plane strain, \(\eta_c\). This lattice distortion changes the V--O--V bond geometry and thereby renormalizes the magnetic exchange interactions. To quantify this effect, we selected several representative biaxial in-plane strain states \((\eta_a,\eta_b)\) and, for each case, computed the strain dependence of the exchange parameters on \(\eta_c\). These exchange constants were then used in parallel-tempering Monte Carlo simulations to determine the corresponding magnetic ground state and ordering temperature.

\begin{figure}[t]
\centering
\includegraphics[width=0.9\textwidth]{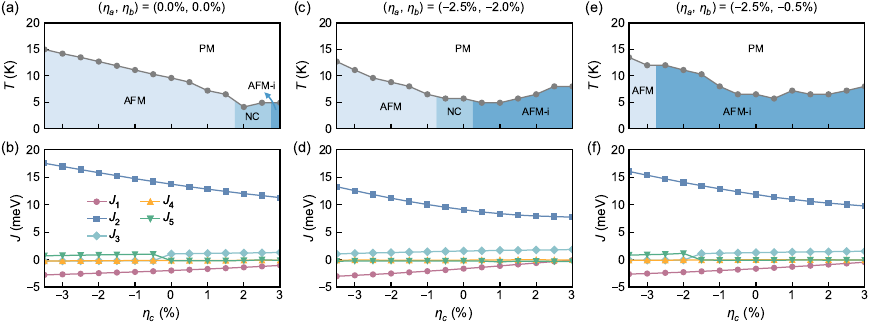}
\caption{Strain-dependent phase diagrams showing the evolution of the transition temperature and magnetic ground state as a function of uniaxial $c$-axis strain ($\eta_c$). The bottom panel shows the corresponding evolution of the exchange interactions under different in-plane strain conditions.
}
\label{order}
\end{figure}

As shown in Fig.~\ref{order}, the out-of-plane strain strongly influences both the magnetic transition temperature and the nature of the magnetic ground state. For the unstrained in-plane lattice (Figs.~\ref{order}a-b), varying \(\eta_c\) primarily tunes the ordering temperature and, at sufficiently large tensile strain, drives a transition from the AFM phase to the NC phase. When compressive in-plane strain is already present (Fig.~\ref{order}c--f), the out-of-plane strain changes the relative strengths of the competing exchange interactions, particularly the balance between \(J_1\) and \(J_2\), and thereby shifts the phase boundaries among the AFM, NC, AFM-i, and paramagnetic regimes. Overall, \(J_2\) remains the dominant exchange channel throughout, but its magnitude varies substantially with \(\eta_c\); at the same time, the weaker couplings evolve systematically and collectively reshape the frustrated magnetic energy landscape.

These results demonstrate strong strain-mediated magnetostructural coupling in VHfO$_4$. Although the electric field does not couple directly to the spins, it modifies the lattice and polarization, which in turn alter the exchange interactions and the relative stability of competing magnetic states. In this way, the ferroelectric distortion provides an indirect but effective route for tuning magnetism.
\note{We further analyzed the magnetic symmetry using the Bilbao Crystallographic Server~\cite{Aroyo06p15}. For the representative AFM structure, the magnetic space group is \Cac{} (BNS No. 9.41, type IV), with magnetic point group \(m.1'\). This magnetic point group is polar and noncentrosymmetric, but it is not compatible with a net ferromagnetic moment. The symmetry-adapted tensor analysis also shows that the linear magnetoelectric tensor \(\alpha_{ij}\) is forbidden, whereas the second-order magnetoelectric tensor \(\alpha_{ijk}\) is allowed. In Bilbao's abbreviated notation, the allowed nonzero second-order components are \(\alpha_{11}\), \(\alpha_{12}\), \(\alpha_{13}\), \(\alpha_{15}\), \(\alpha_{24}\), \(\alpha_{26}\), \(\alpha_{31}\), \(\alpha_{32}\), \(\alpha_{33}\), and \(\alpha_{35}\). Therefore, a direct linear magnetoelectric response is not expected for this magnetic configuration, while nonlinear magnetoelectric response and the strain-mediated mechanism discussed above remain symmetry-allowed.}
This behavior has two important implications. First, at a fixed temperature near the magnetic ordering temperature, an applied electric field could in principle switch magnetic order on or off by moving the system across the boundary between an ordered phase and the paramagnetic state. Second, the same mechanism could drive transitions between distinct ordered phases, such as from AFM to NC, if the system is prepared sufficiently close to the corresponding phase boundary. In practice, the required \(\eta_c\) may be difficult to achieve through piezoelectric deformation alone in a bulk sample. However, this limitation can be mitigated by combining electric-field tuning with appropriate epitaxial pre-strain, which would place the system near a magnetic instability and allow relatively small field-induced lattice distortions to trigger the transition.

\section{Conclusion}

In summary, our first-principles calculations identify stoichiometric VHfO$_4$ as a potential hafnia-derived multiferroic that combines a switchable polar distortion with strain-sensitive magnetic order.
We find that \note{ordered V/Hf layering stabilizes a \Pca-like polar structure that is locally and dynamically stable}, while preserving the characteristic ferroelectric switching pathways of fluorite HfO$_2$.
The calculated polarization and moderate switching barrier indicate that the polar state is not merely structural, but potentially ferroelectrically switchable.

On the magnetic side, the ordered V sublattice introduces competing exchange interactions in the low-symmetry lattice, leading to a nontrivial magnetic energy landscape beyond what can be captured in the primitive cell. By extracting the exchange parameters from DFT and combining them with parallel-tempering Monte Carlo simulations, we show that the zero-strain ground state is antiferromagnetic, with the spin arrangement governed primarily by the competition between the strong ferromagnetic intrachain coupling and weaker antiferromagnetic interchain and interlayer couplings. Under epitaxial strain, this balance changes substantially, giving rise to additional magnetic phases, including a genuinely noncollinear spiral-like state and a predominantly in-plane antiferromagnetic state. These results establish that magnetic frustration in VHfO$_4$ is an intrinsic consequence of the distorted fluorite lattice and can be tuned systematically by strain.

We further show that the exchange interactions are sensitive not only to biaxial epitaxial strain but also to out-of-plane lattice distortions associated with the polar axis. This provides a plausible mechanism for strain-mediated magnetoelectric coupling: an electric field applied along the ferroelectric axis can, in principle, alter the lattice through the converse piezoelectric effect, thereby modifying the exchange couplings, shifting magnetic phase boundaries, and tuning the magnetic transition temperature. \note{Symmetry analysis indicates that the linear magnetoelectric tensor is forbidden for the representative AFM magnetic point group, while a second-order magnetoelectric tensor is allowed.} Although the predicted ordering temperatures remain low and the required structural distortions may be challenging to achieve in bulk form, the calculations suggest that these constraints could be alleviated near epitaxially engineered phase boundaries, where relatively small field-induced strains may be sufficient to trigger magnetic switching.

These results demonstrate a viable design strategy for hafnia-based multiferroics beyond the dilute-doping limit. Rather than relying on randomly distributed magnetic dopants, VHfO$_4$ realizes a dense, ordered magnetic sublattice within a fluorite-derived ferroelectric framework, enabling well-defined exchange pathways and emergent strain-tunable magnetic phases. More broadly, this work shows that ordered transition-metal hafnates can provide a realistic platform for exploring multiferroicity, frustrated magnetism, and electric-field control of magnetic order in materials that remain conceptually connected to CMOS-compatible hafnia.

\begin{acknowledgments}
This work is supported by National Key R\&D Program of China (2021YFA1202100) and National Natural Science Foundation of China (12074319 and 12404114). The computational resource is provided by Westlake HPC Center.
\end{acknowledgments}
\clearpage
\begin{table}[ht]
    \centering
    \caption{Calculated lattice constants (\AA) and relative energies $\Delta E$ (meV/atom) for stoichiometric VHfO$_4$ derived from four representative HfO$_2$ structural polymorphs. The energy of the polar $Pca2_1$-like phase calculated with nonmagnetic is set as the reference. For comparison, the relative energies of the ground-state $P2_1/c$ and metastable $Pbca$ phases in pure HfO$_2$ are $-28$ and $-4$ meV/atom, respectively.}
    \label{str}
    \begin{tabular}{ccccccccc}
        \hline
        \hline
        \specialrule{0em}{1pt}{1pt}
        & {Phase} & a & b & c & $\alpha$ & $\beta$ & $\gamma$ & $\Delta E$\\
        \hline
        \specialrule{0em}{1pt}{1pt}
            FM  & $P2_1/c$      & 4.96 & 5.37 & 4.96 & 96.0 & 90 & 90 & $-201.8$\\
                & $P4_2/nmc$    & 5.09 & 5.09 & 4.91 & 90 & 90 & 90 & $-163.8$\\
                & $Pbcn$        & 5.64 & 4.78 & 5.04 & 90 & 90 & 90 & $-184.1$\\
                & $Pbca$        & 5.11 & 9.74 & 4.97 & 90 & 90 & 90 & $-122.9$\\
                & \Pca-like & 5.37 & 4.96 & 4.96 & 90 & 95.9 & 90 & $-202.0$\\ 
        \specialrule{0em}{1pt}{1pt}
        \hline
        \hline
            AFM & $P2_1/c$      & 4.96 & 5.37 & 4.97 & 96.0 & 90 & 90 & $-200.6$\\
                & $P4_2/nmc$    & 5.09 & 5.09 & 4.90 & 90 & 90 & 90 & $-163.8$\\
                & $Pbcn$        & 5.64 & 4.79 & 5.04 & 90 & 90.5 & 90 & $-182.4$\\
                & $Pbca$      & 5.12 & 9.73 & 4.97 & 90 & 90 & 90 & $-122.4$\\
                & \Pca-like & 5.37 & 4.96 & 4.97 & 90 & 95.8 & 90 & $-200.9$\\ 
        \hline
        \hline
            NM  & $P2_1/c$      & 4.98 & 5.36 & 4.94 & 96.4 & 90 & 90 & $0$\\
                & $P4_2/nmc$    & 5.08 & 5.09 & 4.91 & 90 & 90 & 90 & $44.4$\\
                & $Pbcn$        & 5.66 & 4.75 & 5.06 & 90 & 90.5 & 90 & $13.8$\\
                & $Pbca$        & 5.11 & 9.71 & 4.96 & 90 & 90 & 90 & $59.6$\\
                & \Pca-like & 5.36 & 4.98 & 4.94 & 90 & 96.3 & 90 & $-0.1$\\ 
        \hline
        \hline
    \end{tabular}
\end{table}

\clearpage
\newpage

\clearpage
\newpage

\clearpage
\newpage

\clearpage
\newpage

\clearpage
\newpage

\clearpage
\newpage

\clearpage
\newpage

\clearpage
\newpage

\clearpage
\bibliography{SL}
\end{document}